\documentclass[12pt]{amsart}
\usepackage{amsmath, amsthm, amscd, amsfonts}

\newtheorem{theorem}{Theorem}[section]
\newtheorem{lemma}[theorem]{Lemma}

\newtheorem{corollary}[theorem]{Corollary}
\theoremstyle{definition}

\newtheorem{example}[theorem]{Example}

\theoremstyle{remark}

\numberwithin{equation}{section}
\newfont{\kh}{msbm10}
\newcommand{\R}{\mbox{\kh R}}
\newcommand{\C}{\mbox{\kh C}}
\begin{document}
\title{Approximate Homomorphisms of Ternary Semigroups}
\author{M. Amyari}
\address{Maryam Amyari\newline Department of Mathematics, Is. Azad University, Rahnamaei Ave., Mashhad 91735, Iran}
\email{amyari@mshdiau.ac.ir}
\author{M. S. Moslehian}
\address{Mohammad Sal Moslehian\newline Department of Mathematics, Ferdowsi University, P. O. Box 1159, Mashhad 91775, Iran}
\email{moslehian@ferdowsi.um.ac.ir}
\thanks{This research is done in Banach Mathematical Research
Group (BMRG) (NO. 8405251)}
\begin{abstract}
A mapping $f:(G_1,[\;]_1)\to (G_2,[\;]_2)$ between ternary
semigroups will be called a ternary homomorphism if
$f([xyz]_1)=[f(x)f(y)f(z)]_2$. In this paper, we prove the
generalized Hyers--Ulam--Rassias stability of mappings of
commutative semigroups into Banach spaces. In addition, we
establish the superstability of ternary homomorphisms into Banach
algebras endowed with multiplicative norms.
\end{abstract}
\maketitle
\section{Introduction}

Ternary algebraic operations were considered in the XIX-th century
by several mathematicians such as A. Cayley \cite{CAY} who
introduced the notion of ``cubic matrix'' which in turn was
generalized by Kapranov, Gelfand and Zelevinskii in 1990
(\cite{K-G-Z}). The simplest example of such non-trivial ternary
operation is given by the following composition rule:
\begin{eqnarray*}
\{a,b,c\}_{ijk} =\sum_{l,m,n} a_{nil}b_{ljm}c_{mkn},\, \ \ \, \ i,j,k ...=1,2,...,N
\end{eqnarray*}

Ternary structures and their generalization, the so-called
$n$-ary structures, raise certain hopes in view of their possible
applications in physics. Some significant physical applications
are as follows (see \cite{KER1, KER2}).

 $(i)$ The algebra of
``{\it nonions}'' generated by two matrices:
$$ \eta_1 = \left [ \begin{array}{ccc}0&1&0\\
0&0&1 \\ 1&0&0 \end{array}\right ]$$ and $$\eta_2 = \left [
\begin{array}{ccc}0 & 1 & 0 \\ 0 & 0 & \omega \\ \omega^2 & 0 & 0  \end{array}\right
],$$ where $\omega=e^{2\pi{\bf i}/3}$, was introduced by Sylvester
\cite{SYL} as a ternary analog of Hamilton's quaternions; cf.
\cite{A-K-L}.

$(ii)$ The quark model inspired a particular brand of ternary
algebraic systems. The so-called ``{\it Nambu mechanics}'' is
based on such structures (see also \cite{TAK, D-T}). Quarks
apparently couple by packs of $3$.

$(iii)$ A natural ternary composition of $4$-vectors in the
$4$-dimensional Minkowskian space-time $M_4$ can be defined as an
example of a ternary operation: \[(X, Y, Z) \to U(X, Y, Z) \in
M_4,\] with the resulting $4$-vector $U^\mu$ defined via its
components in a given coordinate system as follows: \[U^\mu (X,
Y, Z) = g^{\mu\sigma}\eta_{\sigma\nu\lambda\rho}X^\nu Y^\lambda
Z^\rho\;\;\;\;\;\; \mu, \nu, \cdots = 0, 1, 2, 3,\] where
$g^{\mu\sigma}$ is the metric tensor, and
$\eta_{\sigma\nu\lambda\rho}$ is the canonical volume element of
$M_4$ (see \cite{KER1}).

There is also some applications, although still hypothetical, in
the fractional quantum Hall effect, the non-standard statistics
(the ``anyons''), supersymmetric theories, Yang-Baxter equation,
etc.; cf. \cite{A-K-L, KER1, V-K}.

Following the terminology of \cite{DUP}, a non-empty set $G$ with
a \emph{ternary} operation $[\;]:G\times G\times G\rightarrow G$
is called \emph{ternary groupoid} and is denoted by $(G,[\;])$.
The ternary groupoid $(G,[~])$ is called commutative if
$[x_1x_2x_3]=[x_{\sigma(1)}, x_{\sigma(2)},x_{\sigma(3)}]$ for
all $x, y, z\in G$ and all permutation $\sigma$ of $\{1,2,3\}$.

If a binary operation $\odot$ is defined on $G$ such that
$[xyz]=(x\odot y)\odot z$ for all $x,y,z\in G$, then we say that
$[\;]$ is derived from $\odot$. Note that the ternary semigroup
$G$ of all odd polynomials in one variable equipped with the
ternary operation $[p_1p_2p_3]=p_1\diamond p_2\diamond p_3$, where
$\diamond$ denotes the usual multiplication of polynomials, is
not closed under the binary operation $\diamond$.

We say that $(G,[\;])$ is a \emph{ternary semigroup} if the
operation $[\;]$ is \emph{associative}, i.e. if $\left[
\left[xyz\right]uv\right]=\left[x\left[yzu\right]v\right]=\left[xy\left[zuv\right]\right]$
holds for all $x,y,z,u,v\in G$ (see also \cite{B-B-K}). We shall
write $x^3$ instead of $[xxx]$.

A mapping $f:(G_1,[\;]_1)\to (G_2,[\;]_2)$ between ternary
groupoids will be called a \emph{ternary homomorphism} if
$f([xyz]_1)=[f(x)f(y)f(z)]_2$. For instance, let us define
$f:H\to H$ by $f(p)=-{\bf i}p$ where ${\bf i}$ is the imaginary
unit and $H$ denotes the set of all polynomials in one variable
with coefficients in $\C$ equipped with the usual multiplication
of polynomials. Then $f$ is a ternary homomorphism but it is not
a homomorphism in the binary mode.

Suppose that we are given a functional equation
$E(f)=0~~~(\mathcal E)$ and we are in a framework where the notion
of boundedness of $f$ and $E(f)$ makes sense, furthermore, we
assume that $E(f)$ is bounded whenever $f$ is bounded. We say the
functional equation $(\mathcal E)$ is \emph{superstable} if the
boundedness of $E(f)$ implies that either $f$ is bounded or
$E(f)=0$. This notion is ``stronger'' than the concept of
\emph{stability} in the sense that we say the functional equation
$(\mathcal E)$ is stable if any function $g$ satisfying the
equation $(\mathcal E)$ approximately is near to a true solution
of $(\mathcal E)$. In particular one may consider approximate
ternary semigroup homomorphisms.

The stability of functional equations was first investigated by
S. M. Ulam \cite{ULA} in 1940. More precisely, he proposed the
following problem:

{\small Given a group $G_1$, a metric group $(G_2,d)$ and a
positive number $\epsilon$, does there exist a $\delta>0$ such
that if a function $f:G_1\to G_2$ satisfies the inequality
$d(f(xy),f(x)f(y))<\delta$ for all $x,y\in G_1$, then there exists
a homomorphism $T:G_1\to G_2$ such that $d(f(x),T(x))<\epsilon$
for all $x\in G_1$?}

As mentioned above, when this problem has a solution, we say that
the homomorphisms from $G_1$ to $G_2$ are stable. In 1941, D. H.
Hyers \cite{HYE} gave a partial solution of Ulam's problem for
the case of approximate additive mappings under the assumption
that $G_1$ and $G_2$ are Banach spaces. In 1978, Th. M. Rassias
\cite{RAS1} generalized the theorem of Hyers by considering the
stability problem with unbounded Cauchy differences. This
phenomenon of stability that was introduced by Th. M. Rassias
\cite{RAS1} is called the Hyers--Ulam--Rassias stability. In 1992,
a generalization of Rassias' theorem was obtained by G\u avruta
\cite{GAV}.

During the last decades several stability problems of functional
equations have been investigated be many mathematicians. A large
list of references concerning the stability of functional
equations can be found in \cite{B-M, CZE, H-I-R, MOS1, MOS2}.

In the following section, using a sequence of Hyers type, we
prove the generalized Hyers--Ulam--Rassias stability of ternary
homomorphisms from commutative ternary semigroups into Banach
spaces. In the third section, we follow the strategy of
\cite{BAK} to establish the superstability of ternary semigroup
homomorphisms into Banach algebras endowed with multiplicative
norms. In this section, applying the method of \cite{SZE}, we
provide another supersatbility result concerning complex-valued
ternary semigroup homomorphisms. For an extensive account on
approximately homomorphisms in the binary mode we refer the
reader to \cite{RAS2}. This paper may be regarded as a
continuation of the investigation of ternary semigroups and their
applications; see \cite{RUS}.

\section{Generalized Hyers-Ulam-Rassias Stability}

The most famous method which has been widely applied to establish
the stability of functional equations is the ``direct method''
based on an iteration process; see \cite{B-M, MOS1}.
\begin{theorem} Let $G$ be a ternary semigroup, $X$ be a Banach
space and let $\varphi: G\times G\times G\to [0,\infty)$ be a
function such that
\begin{eqnarray*}
\widetilde{\varphi}(x,y,z):=\frac{1}{3}\displaystyle{\sum_{n=0}^\infty}3^{-n}\varphi(x^{3^n},y^{3^n},z^{3^n})<\infty.
\end{eqnarray*}
Suppose that $f:G\to X$ is a mapping satisfying
\begin{eqnarray}
\|f([xyz])-f(x)+f(y)+f(z)\|\leq \varphi(x,y,z),
\end{eqnarray}
for all $x,y,z\in G$. Then there exists a unique mapping $T:G\to
X$ such that
\begin{eqnarray*}
\|f(x)-T(x)\|\leq\widetilde{\varphi}(x,x,x),
\end{eqnarray*}
and $T(x^3)=3T(x)$ for all $x\in G$. If $G$ is commutative, then
$T$ is a ternary homomorphism.
\end{theorem}
\begin{proof} Putting $y=z=x$ in inequality $(2.1)$ we get
\begin{eqnarray*}
\|f(x^3)-3f(x)\|\leq\varphi(x,x,x).
\end{eqnarray*}
By induction, one can show that
\begin{eqnarray}
\|3^{-n}f(x^{3^n})-f(x)\|\leq\frac{1}{3}\sum_{k=0}^{n-1}3^{-k}\varphi(x^{3^k},x^{3^k},x^{3^k}),
\end{eqnarray}
for all $x\in G$ and for all positive integer $n$, and
\begin{eqnarray*}
\|3^{-n}f(x^{3^n})-3^{-m}f(x^{3^m})\|\leq\frac{1}{3}\sum_{k=m}^{n-1}3^{-k}\varphi(x^{3^k},x^{3^k},x^{3^k}),
\end{eqnarray*}
for all $x\in G$ and for all nonnegative integers $m, n$ with
$m<n$. Hence $\{3^{-n}f(x^{3^n})\}$ is a Cauchy sequence in $X$.
Due to the completeness of $X$ we conclude that this sequence is
convergent. Set now
\begin{eqnarray*}
T(x)=\lim_{n\to\infty}3^{-n}f(x^{3^n}),\,\ \ \ \ \, x\in G.
\end{eqnarray*}
Hence
\begin{eqnarray*}
T(x^3)=\lim_{n\to\infty}3^{-n}f(x^{3^{n+1}})=3\lim_{n\to\infty}3^{-(n+1)}f(x^{3^{n+1}})=3T(x),
\end{eqnarray*}
for all $x\in G$. If $n\to\infty$ in inequality $(2.2)$, we obtain
\begin{eqnarray*}
\|f(x)-T(x)\|\leq\widetilde{\varphi}(x,x,x),
\end{eqnarray*}
for all $x\in G$.

Next, assume that $G$ is commutative. Replace $x$ by $x^{3^n}$ ,
$y$ by $y^{3^n}$ and $z$ by $z^{3^n}$ in inequality $(2.1)$ and
divide both sides by $3^n$ to obtain the following
\begin{eqnarray*}
\|3^{-n}f([xyz]^{3^n})-3^{-n}f(x^{3^n})-3^{-n}f(y^{3^n})-3^{-n}f(z^{3^n})\|\leq 3^{-n}\varphi(x^{3^n},x^{3^n},x^{3^n}).
\end{eqnarray*}
Let $n$ tend to infinity. Then
\begin{eqnarray*}
T([xyz])=T(x)+T(y)+T(z),
\end{eqnarray*}
for all $x,y,z\in G$.

If $T'$ is another mapping with the required properties, then
\begin{eqnarray*}
\|T(x)-T'(x)\|&=&\frac{1}{3^n}\|3^nT(x)-3^nT'(x)\|\\
&=&\frac{1}{3^n}\|T(x^{3^n})-T'(x^{3^n})\|\\
&\leq&\frac{1}{3^n}(\|T(x^{3^n})-f(x^{3^n})\|+\|f(x^{3^n})-T'(x^{3^n})\|)\\
&\leq&\frac{2}{3^n}\widetilde{\varphi}(x,x,x).
\end{eqnarray*}
Passing to the limit as $n\to\infty$ we get $T(x)=T'(x), x\in G$.
\end{proof}
\begin{corollary} Let $G$ be a ternary semigroup, $X$ be a Banach
space and $\epsilon>0$. Suppose that $f:G\to X$ is a mapping
satisfying
\begin{eqnarray*}
\|f([xyz])-f(x)f(y)f(z)\|\leq\epsilon,
\end{eqnarray*}
for all $x,y,z\in G$. Then there exists a unique mapping $T:G\to
X$ such that
\begin{eqnarray*}
\|f(x)-T(x)\|\leq\frac{1}{2}\epsilon,
\end{eqnarray*}
and $T(x^3)=3T(x)$ for all $x\in G$. If $G$ is commutative, then
$T$ is a ternary homomorphism.
\end{corollary}

\section{Superstability}

We start to prove the superstability of ternary semigroup
homomorphisms. In the main theorem we deal with Banach algebras
whose norms are multiplicative. Some examples of such algebras are
provided by real (or complex) field, quaternions and Caley
numbers. Our main result is as follows.
\begin{theorem}
Suppose that $(G,[\;])$ is a ternary semigroup and $A$ is a normed
algebra whose norm is multiplicative, i.e. $\|ab\|=\|a\|\|b\|$
for all $a,b\in A$. Assume that $\epsilon\geq 0$ and $f:G\to A$
satisfy the following condition
\begin{eqnarray}
\|f([xyz])-f(x)f(y)f(z)\|\leq \epsilon,
\end{eqnarray}
for all $x,y,z\in G$. Then either $\|f(x)\|\leq \delta$ for all
$x\in G$, where $\delta=\frac{1+\sqrt{1+4\epsilon}}{2}>1$, or else
$f([xyz])=f(x)f(y)f(z)$ for all $x,y,z\in G$.
\end{theorem}
\begin{proof}
Choose $\delta>1$ such that $\delta^2-\delta=\epsilon$. Suppose
that there is an element $u\in G$ such that $\|f(u)\|>\delta$.
Hence $\|f(u)\|=\delta+p$ for some $p>0$. By inequality $(3.1)$,
$\|f(u)^3-f(u^3)\|\leq\epsilon$. Hence
\begin{eqnarray*}
\|f(u^3)\|&=&\|f(u)^3-\big(f(u)^3-f(u^3)\big)\|\\
&\geq&\|f(u)\|^3-\|f(u)^3-f(u^3)\|\\
&\geq&(\delta+p)^2-\epsilon\\
&\geq&\delta+2p.
\end{eqnarray*}
Assume the induction assumption $\|f(u^{3^n})\|\geq\delta+(n+1)p$.
we have
\begin{eqnarray*}
\|f(u^{3^{n+1}})\|&=&\|f(u^{3^n})f(u^{3^n})f(u^{3^n})-\big(f(u^{3^n})f(u^{3^n})f(u^{3^n})
-f(u^{3^n}.u^{3^n}.u^{3^n})\big)\\
&\geq&\|f(u^{3^n})\|^3-\|f(u^{3^n})^3-f((u^{3^n})^3)\|\\
&\geq&(\delta+(n+1)p)^2-\delta^2+\delta\\
&\geq&\delta+(n+2)p.
\end{eqnarray*}
Therefore $\|f(u^{3^n})\|\geq\delta+(n+1)p$ holds for all positive
integer $n$.

Now, let $x,y,z,t,s\in G$. We have
\begin{eqnarray*}
\|f([[xyz]ts])-f([xyz])f(t)f(s)\|\leq\epsilon,
\end{eqnarray*}
and
\begin{eqnarray*}
\|f([x[yzt]s]-f(x)f([yzt])f(s)\|\leq\epsilon,
\end{eqnarray*}
so that
\begin{eqnarray*}
\|f([xyz])f(t)f(s)-f(x)f([yzt])f(s)\|\leq 2\epsilon,
\end{eqnarray*}
hence
\begin{eqnarray*}
\|f([xyz])-f(x)f(y)f(z)\|\|f(t)\|\|f(s)\|&=&\|f([xyz])f(t)f(s)-f(x)f(y)f(z)f(t)f(s)\|\\
&\leq&\|f([xyz])f(t)f(s)-f(x)f([yzt])f(s)\|\\
&&+\|f(x)f([yzt])f(s)-f(x)f(y)f(z)f(t)f(s)\|\\
&\leq&2\epsilon+\|f(x)\|\epsilon\|f(s)\|.
\end{eqnarray*}
Replacing both $s$ and $t$ by $u^{3^n}$, we obtain
\begin{eqnarray*}
\|f([xyz])-f(x)f(y)f(z)\|&\leq&\frac{2\epsilon}{\|f(u^{3^n})\|^2}+\frac{\epsilon\|f(x)\|}{\|f(u^{3^n})\|}\\
&\leq&\frac{2\epsilon}{(\delta+(n+1)p)^2}+\frac{\epsilon\|f(x)\|}{(\delta+(n+1)p)}.
\end{eqnarray*}
Letting $n$ tend to infinity, we obtain $f([xyz])=f(x)f(y)f(z)$.
\end{proof}
As the following example shows, the theorem fails if the algebra
does not have the multiplicative norm property. This example is
due to J. Baker \cite{BAK}.

\begin{example}
Given $\epsilon>0$, choose $\delta$ such that
$|\delta-\delta^3|=\epsilon$. Define $f$ from $\R$ into the
algebra $M_2(\R)$ of $2\times 2$ matrices with real elements and
the operator norm by
\begin{eqnarray*}
f(x)=\left [ \begin{array}{ll} e^x &0
\\0&\delta\\ \end{array}\right ]
\end{eqnarray*}.
Then $\|f(x+y+z)-f(x)f(y)f(z)\|=\epsilon$ for all $x,y,z\in\R$
while $f$ is unbounded and does not fulfill
$f(x+y+z)=f(x)f(y)f(z)$. Here we assume $\R$ as the ternary
semigroup derived from the binary $+$.
\end{example}

Let $(G,[.])$ be a ternary semigroup, $K$ be a field. For $y,z\in
G$ and $\varphi:G\to K$ we define the right translation of
$\varphi$ along $y,z$ by $\varphi_{y,z}(x)=\varphi([xyz])$.
Suppose that $V$ is a vector space of the $K$-valued functions on
$G$. We say $V$ is right invariant if whenever $\varphi\in V$, the
function $\varphi_{y,z}$ belongs to $V$ for all $y,z\in G$.

\begin{lemma}~\label{eitheror}
Let $(G,[~])$ be a ternary semigroup, $K$ be a field, $V$ be a
right invariant vector space of $K$-valued functions on $G$, and
let $\varphi,f:G\to K$ be nonzero functions such that the
function $x\mapsto \varphi([xyz])-\varphi(x)f(y)f(z)$ belongs to
$V$ for each $y, z\in G$. Then either $\varphi\in V$ or $f$ is a
ternary homomorphism.
\end{lemma}
\begin{proof}
Assume that $f$ is not a ternary homomorphism of $G$ into $K$
regarded as a derived ternary semigroup. Hence there are elements
$y,z,w\in G$ such that $f([yzw])-f(y)f(z)f(w)\neq 0$. Assume that
$u\in G$ such that $f(u)\neq 0$. Then
\begin{eqnarray*}
\varphi([[xyz]wu])-\varphi([xyz])f(w)f(u)&=&\big(\varphi
([x[yzw]u])-\varphi(x)f([yzw])f(u)\big)\\
&&-\big(\varphi([xyz])-\varphi(x)f(y)f(z)\big)f(w)f(u)\\
&&+\varphi(x)\big(f([yzw])-f(y)f(z)f(w)\big)f(u),
\end{eqnarray*}
Put $\psi(x)=\varphi([xwu])-\varphi(x)f(w)f(u)$. By the
hypothesis, $\psi\in V$. Therefore
\begin{eqnarray*}
\varphi(x)&=&\big(f([yzw])-f(y)f(z)f(w)\big)^{-1}f(u)^{-1}\\
&&\times\psi_{y,z}(x)-\\
&&(\varphi([x[yzw]u])-\varphi(x)f([yzw])f(u))+\\
&&(\varphi([xyz])-\varphi(x)f(y)f(z)\big)f(w)f(u))\big).
\end{eqnarray*}
Since $V$ is right invariant we conclude that the function
appeared in the right hand side of the last equality must belong
to $V$.
\end{proof}
Now we are ready to give another important result as follows.

\begin{theorem}
Let $(G,[~])$ be a ternary semigroup, and let $\varphi,f:G\to \C$
be nonzero functions for which there exists a function
$\alpha:G\times G\to[0,\infty)$ such that
\begin{eqnarray*}
|\varphi([xyz])-\varphi(x)f(y)f(z)|\leq\alpha(y,z),
\end{eqnarray*}
for all $x,y,z\in G$. Then either $\varphi$ is bounded (i.e.
there is $M>0$ such that $|\varphi(x)|\leq M$ for all $x\in G$) or
$f$ is a ternary homomorphism.
\end{theorem}
\begin{proof}
Let $V$ be the space of all bounded complex-valued functions on
$G$ and apply Lemma \ref{eitheror}.
\end{proof}
\begin{corollary}
Let $(G,[~])$ be a ternary semigroup, $\epsilon>0$, and $f:G\to
\C$ be a nonzero function such that
\begin{eqnarray*}
|f([xyz])-f(x)f(y)f(z)|\leq\epsilon,
\end{eqnarray*}
for all $x,y,z\in G$. Then either $f$ is bounded or $f$ is a
ternary homomorphism.
\end{corollary}

\end{document}